\documentclass[showpacs,aps,amssymb,floatfix,prd,preprintnumbers]{revtex4}
\usepackage{epstopdf}
\usepackage{capt-of}
\usepackage{graphicx}  
\usepackage{dcolumn}   
\usepackage{caption} 
\usepackage{bm}
\usepackage{amsmath}
\usepackage[font=scriptsize]{caption}
\usepackage{subcaption}
\usepackage{multirow}
\usepackage[colorlinks]{hyperref}
\usepackage{xcolor}
\usepackage{orcidlink}
\begin{document}
\input epsf.tex
\title{\bf Late-Time Dynamics in Accelerating Nonlinear $f(T)$ Gravity via Hybrid Scale Factor : Bayesian Analysis and Cosmological implications}

\author{Suraj Kumar Behera \orcidlink{0009-0009-9294-460X}}
\email{skbehera.researches@gmail.com}
\affiliation{Department of Mathematics, School of Advanced Sciences, VIT-AP University, Beside AP Secretariat, Amaravati, 522241, Andhra Pradesh, India.}

\author{Pratik P. Ray \orcidlink{0000-0003-2304-0323}}
\email{pratik.chika9876@gmail.com} 
\email[]{pratik.ray@vitap.ac.in}
\affiliation{Department of Mathematics, School of Advanced Sciences, VIT-AP University, Beside AP Secretariat, Amaravati, 522241, Andhra Pradesh, India}
\affiliation{Pacif Institute of Cosmology and Selfology (PICS), Sagara, Sambalpur 768224, Odisha, India.}
\begin{abstract}
\begin{center}
{\textbf{Abstract}}
\end{center} 

This study offers a comprehensive reconstruction of $f(T)$ gravity model with three distinct non-linear as well as novel forms employing a hybrid scale factor to depict the expansion history of the universe. 
We investigate a physically motivated cosmological parametrization through this hybrid scale factor that can explain a smooth transition from an early decelerated epoch to the present accelerated evolution. This scale factor is then used to derive the corresponding model parameters. These parameters are rigorously constrained using the Monte Carlo Markov Chain (MCMC) analysis with the help of Bayesian statistics and incorporating late-time observations from Hubble, Baryon Acoustic Oscillations (BAO) and Patheon+SH0ES datasets. The investigation of dynamical parameters such as the equation of state parameter and cosmological parameters indicates alignment with an accelerated expansion phase in both the present and late time epochs. Validation is conducted by assessing the energy conditions, verifying the feasibility of the model forms with particular emphasis on the violation of the strong energy condition that indicates dark energy dominance in modified gravity scenarios. This investigation has been instrumental in determining models that remain consistent with cosmological observations and theoretical requirements. The reconstructed forms of the model effectively mimic $\Lambda$CDM at late times, providing significant insights into possible extensions of general relativity and bolstering $f(T)$ gravity theory as a robust explanation for cosmic acceleration.
\end{abstract}

\maketitle

\section{Introduction}

The observation of accelerated cosmic expansion signifies a landmark development in modern cosmology, fundamentally reshaping our comprehension of large-scale structure and energy composition of the universe. Type Ia Supernovae were the initial to observe this phenomenon \cite{Riess_1998,Perlmutter_1999}, and precise measurements of the Cosmic Microwave Background (CMB) \cite{2007ApJS..170..377S,Komatsu_2009} and Baryon Acoustic Oscillations (BAO) \cite{Eisenstein_2005,10.1111/j.1365-2966.2009.15812.x} verified it later. A more comprehensive inquiry into the primary causes of the accelerated expansion has been stimulated by accumulating the observational evidence, which has demonstrated it as an accepted feature of our universe \cite{WEINBERG201387}. This unexplained phenomena has been ascribed to an enigmatic component that is dark energy (DE), which constitutes a substantial amount of the overall energy budget of the universe \cite{doi:10.1142/S0219887807001928}. In the beginning, a cosmological constant $\Lambda$ is put forward as the most fundamental type of DE in the standard cosmological model, $\Lambda$CDM model, which is used as the framework \cite{riess2019large,NAJERA2021100889,Junior_2016}. Despite its achievements with the large-scale structure of the universe \cite{10.1093/mnras/staa3336}, $\Lambda$CDM has theoretical problems. These include the coincidence problem, which arises question that the near-equality of matter and DE densities in the current epoch \cite{Cai_2005}, and the cosmological constant problem, in which quantum field theory predicts a vacuum energy that is substantially greater than what is observed \cite{RevModPhys.61.1}.\\

Because of these constraints, a number of modified theories of gravity have been introduced to explain the acceleration phenomena of the universe without relying on a cosmological constant \cite{RevModPhys.82.451,PhysRevD.98.044048}
. Teleparallel gravity is one of the more appealing types of modified theories. Instead of implementing the traditional spacetime metric $g_{\mu \nu}$ to convey gravitaional effects, this gravity uses a tetrad field $e^i_{\mu}$. Torsion, which is generated by the tetrads, is enough to represent the gravity. This approach led to the establishment of $f(T)$ gravity, where torsion takes the place of curvature to produce a flat spacetime framework. and the Teleparallel equivalent of general relativity (TEGR) \cite{rovere2025covariantfractonsweitzenbocktorsion,PhysRevD.19.3524,osti_4688687}. $f(T)$ gravity has a major advantage, which is its second-order field equations rather than the fourth-order in $f(R)$. Without using DE, this theory has been excellent at describing late-time cosmic acceleration, thereby making it an excellent contender to tackle unanswered questions in cosmology. Tamanini and Boehmer \cite{PhysRevD.86.044009} examined the dynamics of $f(T)$ gravity models, paying special attention to their cosmological implications and stability. The literature \cite{Bamba_2011} discusses an interacting cosmic fluid with $f(T)$ gravity that is characterized by a generalized Chaplygin gas with viscosity. In their study of different aspects of $f(T)$ gravity, Dent et al. \cite{PhysRevD.83.023508} explored how minor deviations from general relativity might show up in cosmological measurements. In the case of a charged black hole, $f(T)$ gravity provides an analytical solution \cite{Capozziello_2013}. Geng et al. \cite{Geng_2012} analyzed the stability of cosmological models while focusing on the dynamics of $f(T)$ gravity in the presence of a perfect fluid. Also, some other investigations in the framework of $f(T)$ gravity are large-scale structure \cite{PhysRevD.83.104017}, accelerated expansion \cite{Duchaniya_2022}, finite time singularity \cite{PhysRevD.85.104036,Bamba_2012}.\\

The cosmic transition from early deceleration to present accelerated expansion has been investigated with a hybrid scale factor, which combines exponential and power-law behaviours to model a smooth evolution of the universe. Mishra and Tripathy\cite{doi:10.1142/S0217732315501758} developed an anisotropic dark energy model within Bianchi type V spacetime, employing a hybrid scale factor to characterize the comsic transition from early deceleration to late-time acceleration. An anisotropy dark energy model in Bianchi type V spacetime with two non-interacting fluids-cosmic strings and dark energy-is developed using a hybrid scale factor in the Ref.\cite{doi:10.1142/S0219887817501249}. Mishra, Ray and Pacif\cite{Mishra_2017} examined anisotropic dark energy cosmological models within Bianchi type V spacetime by considering three genral forms of the scale factor,  including a hybrid one. Mishra \textit{et al.}\cite{doi:10.1142/S0217732318500529} examined Bianchi type V anisotropic cosmologies inside the extended $f(R,T)$ gravity framework. Narawade, Koussour, and Mishra\cite{Narawade_2023} explored an anisotropic cosmological model in the framework of $f(Q,T)$ gravity, employing a hybrid scale factor with model parameters constrained by Hubble, BAO , and Pantheon datasets.\\

In this study, we investigate the dynamics and behaviour of three well-motivated non-linear forms of $f(T)$ such as $f(T)=T-\frac{a}{(-T/6)^n}+b(\frac{-T}{6})^m$, $f(T)=T-aT_0[(1+\frac{T^2}{T_0^2})^{-n}-1]$, and $f(T)=T+aT_0\frac{(T^2/T_0^2)^n}{1+(T^2/T_0^2)^n}$ considering the hybrid scale factor with constraints on its parameters through the Monte Carlo Markov Chain (MCMC) approach using Bayesian statistics. The evolutionary behavior is verified with the equation of state parameter, and also, the energy conditions are examined. This paper follows a format like: in Sec.\textbf{II}, the field equations of $f(T)$ are formulated, and also, the dynamical parameters and the equation of state (EoS) parameter ($\omega$) are derived. Sec.\textbf{III} discusses the hybrid scale factor and the Hubble model derived from it. Model parameters are constrained by fitting with observational datasets through the MCMC algorithm in sec.\textbf{IV}. The dynamics of the newly reconstructed $f(T)$ models are investigated in sec.\textbf{V} by plotting dynamical parameters. The energy conditions of the models are examined in sec.\textbf{VI}. Finally, in the sec.\textbf{VII}, the conclusions of this study have been discussed, followed by the future scope of research.

\section{Mathematical framework for field equations}

In the Teleparallel formulation of gravity, by employing the tetrads as the fundamental dynamical variables, instead of the metric tensor, General Relativity (GR) can be reconstructed in an equivalent way \cite{einstein2005riemann}. At each point in spacetime, the tetrad field creates a basis ${e_{\mu}(x)}$, where $\mu$ varies from $0$ to $3$, representing four linearly independent vectors. It will be feasible to expand each tetrad vector $e_{\mu}$ in the coordinate basis to get the tetrad components $e_{\mu}^a$. Hence, the orthogonality relation between the tetrads and the spacetime metric is expressed as

\begin{equation}
    g_{\mu \nu} = \eta_{ab}e^a_{\mu}e^b_{\nu}
\end{equation}

where $g_{\mu \nu}$ is the metric tensor and $\eta_{ab} = diag(1,-1,-1,-1)$ represents the Minkowski metric. The inverse relation takes the form

\begin{equation}
    E^{\mu}_ae^b_{\mu}=\delta^b_a
\end{equation}

and the tetrad is $e=det[e^a_{\mu}=\sqrt{-g}]$. \\

The witzenb{\"o}ck connection \cite{aldrovandi2012Teleparallel} is defined as $\Gamma^{\sigma}_{\ \ \mu \nu} \equiv\ E^{\sigma}_{a}\partial_{\mu}e^{a}_{\nu}$ which leads to a non-vanishing torsion tensor

\begin{equation}
   T^{\sigma}_{\ \ \mu \nu} \equiv\ {\Gamma}^{\sigma}_{\ \nu \mu} - {\Gamma}^{\sigma}_{\ \mu \nu} = E^{\sigma}_{a}\partial_{\mu}e^{a}_{\nu} - E^{\sigma}_{a}\partial_{\nu}e^{a}_{\mu} 
\end{equation}

This allows us to achieve the contorsion tensor as 

\begin{equation}
     K^{\mu \nu}_{\ \ \ \sigma} = \frac{1}{2}(T^{\ \ \mu \nu}_{\sigma} + T^{\nu \mu}_{\ \ \ \sigma} - T^{\mu \nu}_{\ \ \ \sigma})
\end{equation}

The torsion scalar is then determined to be

\begin{equation}
    T \equiv\ S^{\ \ \mu \nu}_{\sigma}T^{\sigma}_{\ \ \mu \nu} 
\end{equation}

where the superpotential tensor is defined as

\begin{equation}
    S^{\ \ \mu \nu}_{\sigma} = \frac{1}{2}[K^{\mu \nu}_{\ \ \ \sigma} + \delta^{\mu}_{\sigma}T^{\alpha \nu}_{\ \ \ \alpha} - \delta^{\nu}_{\sigma}T^{\alpha \mu}_{\ \ \ \alpha}]
\end{equation}
 This ensures that the GR and TEGR actions will produce identical field equations.\\

In the framework of TEGR, the gravitational Lagrangian depends linearly on the torsion scalar $T$. This is generalized by the $f(T)$ gravity framework, which substitutes $f(T)$ for $T$. The formula for the action \cite{PhysRevD.79.124019,Myrzakulov:2011} is defined by

\begin{equation}\label{eq:7}
    A= \frac{1}{16 \pi G} \int d^4xe\left[f\left(T\right)+\mathcal{L}_m\right]
\end{equation}

Here, $G$ is the gravitational constant and $\mathcal{L}_m$ is the total matter Lagrangian density. We use natural units throughout this study by setting  $8\pi G = \kappa^2 = c = 1$.\\ 

Taking the variation of the action \eqref{eq:7} with regard to the tetrad fields leads to the corresponding formula of modified field equations, mentioned as,

\begin{equation} \label{field equation formula}
    S^{\ \mu \nu}_{\rho} \partial_{\mu}(T) f_{TT}+\frac{1}{4}E^{\nu}_a f(T) + \left[ e^{-1} \partial_{\mu} \left(e S^{\ \mu \nu}_{\rho} \right)-E^{\sigma}_a T^{\rho}_{\mu \sigma} S^{\nu \mu}_{\rho} \right]f_T = 4 \pi G E^{\rho}_a \mathcal{T}^{\nu}_{\rho}
\end{equation}

where, $f_{T}$ and $f_{TT}$ are first and second derivatives of $f(T)$ with respect to $T$ respectively. Also, $\mathcal{T}^{\ \nu}_{\rho}$ indicates the total matter energy-momentum tensor of perfect fluid in the context.\\

 We notice that the local Lorentz invariance issue may occur in the pure tetrad formulation of generic $f(T)$ gravity. So, here, we consider a spatially flat Friedman-Lema$\hat{i}$tre-Robertson-Walker (FLRW) universe for cosmological applications, which is expressed by the line elements

\begin{equation}
    ds^2 = dt^2-a^2 (t)(dx^2 + dy^2 + dz^2)
\end{equation}

where, $a(t)$ is the scale factor. Moreover, we choose the diagonal tetrad $(1,a(t),a(t),a(t))$, which reproduces the cosmological dynamics without introducing spurious anisotropic or non-physical additional degrees of freedom because of the strong symmetry of the FLRW spacetime.
Hence, it offers a consistent cosmological description for the spatially flat FLRW background. Substituting these components, into \eqref{field equation formula} along with a perfect fluid treatment, the Friedman equations modified for $f(T)$ gravity are obtained as

\begin{equation} \label{Friedmann 1}
    12 H^2 f_T + f = 2 \kappa^2 \rho,
\end{equation}
\begin{equation} \label{Friedmann 2}
48 H^2 \dot{H} f_{TT} - \left(12 H^2 + 4 \dot{H} \right) f_T-f = 2 \kappa^2 p
\end{equation}

Here onward from this field equations we use $f=f(T)$, $f_T$ and $f_{TT}$ are first and second derivatives of $f(T)$ with respect to $T$ and $H \equiv \frac{\dot{a}}{a}$. $\rho$ and $p$ represent the energy density and pressure, respectively. We can derive the energy density and pressure for DE sector by comparing eqs. \eqref{Friedmann 1} and \eqref{Friedmann 2} with Friedmann equations of GR,

\begin{equation}
    3H^2 = \kappa^2 (\rho + \rho_{de}),
\end{equation}

\begin{equation}
    3H^2 + 2 \dot{H} = - \kappa^2 (p+p_{de})
\end{equation}

In this context of $f(T)$ gravity, the effective DE density and pressure are given by

\begin{equation}\label{eq:12}
    \rho_{de} \equiv \frac{1}{16 \pi G} [-12 H^2 f_T -f - T]
\end{equation}
\begin{equation}
    p_{de} \equiv -\frac{1}{16 \pi G} [48 \dot{H} H^2f_{TT} - 4 \dot{H} f_T + 4 \dot{H}] - \rho_{de}
\end{equation}

where we use the relation $T=-6H^2$. The effective EoS parameter for the DE sector is thus expressed as

\begin{equation}\label{eq:14}
    \omega_{de} = -1 - \ \frac{48 \dot{H} H^2f_{TT} - 4 \dot{H} f_T + 4 \dot{H}}{-12 H^2 f_T -f - T}
\end{equation}

A time-dependent EoS parameter is crucial for investigating the accelerated expansion of the universe because it provides a more flexible parametrization with regard to the redshift. To make this analysis simpler, we use a hybrid scale factor in the section that follows. 

\section{ Parametrization of the hybrid scale factor}

Some observations of the present universe indicate that the power-law expansion and the de-sitter solution emerge as appropriate choices for the scale factor. Both the power-law and exponential forms of the scale factor exhibit a constant deceleration parameter. In this study, we examine a certain scale factor, namely the hybrid scale factor \cite{doi:10.1142/S0217732315501758} that inherently results in a constant deceleration parameter at late times. The expression for the hybrid scale factor can be presented as

\begin{equation}
    a(t) = e^{\lambda t}t^{\beta}
\end{equation}

where $\lambda$ and $\beta$ are positive constants. This form blends two different evolutionary behaviors: a power-law expansion with $t^{\beta}$ and exponential expansion with $e^{\lambda t}$. Early cosmic dynamics are dominated by the power-law term, whereas later eras are dominated by the exponential term. The hybrid scale factor eventually evolves to a stage in which the exponential component governs the dynamics, reflecting the smooth transition from an early decelerated phase to present and late-time acceleration period, and precisely simulating the accelerated expansion of the universe. The adopted hybrid scale factor has significant physical motivations and is not introduced arbitrarily to produce an artificial Hubble evolution. In particular, it combines exponential and power-law expansion tendencies, enabling a unified description of different cosmological phases under a single framework.\\

Additionally, this formulation elegantly connects multiple cosmological phases: the standard exponential (de-sitter) expansion is recovered for $\beta = 0$ and the model reduces to a pure power-law expansion for $\lambda = 0$. Thus, the hybrid scale factor delivers a cohesive explanation of the evolution of the universe by providing a natural mechanism to describe a cosmic shift from an early decelerated phase to a late-time accelerated period. The Hubble parameter derived from this model is given by

\begin{equation}
    H(z) = \lambda +\frac{\lambda }{W\left(\frac{\lambda  \left(\frac{1}{z+1}\right)^{1/\beta }}{\beta }\right)}
\end{equation}

where $W\left(\frac{\lambda  \left(\frac{1}{z+1}\right)^{1/\beta }}{\beta }\right) e^{W\left(\frac{\lambda  \left(\frac{1}{z+1}\right)^{1/\beta }}{\beta }\right)} = \frac{\lambda  \left(\frac{1}{z+1}\right)^{1/\beta }}{\beta } $ and the deceleration parameter can be derived as

\begin{equation}
    q(z) = \frac{1}{\beta  \left(W\left(\frac{\lambda  \left(\frac{1}{z+1}\right)^{1/\beta }}{\beta }\right)+1\right)^2}-1
\end{equation}

Both the Hubble and deceleration parameters are in the form of redshift. So, their parameters can be easily constrained by fitting them with various observational datasets.

\section{Observational Data-Driven Analysis}

This section highlights observational cosmology, which is crucial for the establishment of precise and coherent cosmological models. To accomplish this, it is important to constrain the model parameters, specifically $\lambda$  and $\beta$ through rigorous examination of observational data.\\

This investigation utilizes various observational datasets, including Hubble $(H(z))$, Baryon Acoustic Oscillation (BAO) and Pantheon+SH0ES, to assess the validity of our chosen model. Through statistical analysis of these datasets, we find the best-fit model parameters and their associated confidence intervals, providing robust observational constraints on the model. 

\subsection{$H(z)$ Dataset}

Early-type galaxies are used to estimate their differential evolutionary rates in order to obtain the Hubble parameter measurements. The Cosmic Chronometer (CC) strategy is the process by which these measurements are collected. The available data span a redshift range $ 0.07 \leq z \leq 1.965 $ \cite{Zhang_2014,PhysRevD.71.123001,Moresco_2012,moresco20166,Stern_2010,10.1093/mnras/stx301,10.1093/mnrasl/slv037,Borghi_2022}. The $\chi^2$ estimation for this 32-point dataset can be calculated as 

\begin{equation}
    \chi^2_{H(z)} = \sum_{i=1}^{32} \left[\frac{\left(H_\text{th}(z_i, \theta) - H_\text{obs}(z_i)\right)^2}{\sigma_H^2(z_i)}\right],
\end{equation}

Here $H_\text{th}(z_i, \theta)$ symbolises the theoretical Hubble parameter determined at redshift $z_i$ for a specific collection of cosmological parameters $\theta$, while $H_\text{obs}(z_i)$ reflects the associated observed values. The quantity $\sigma_H^2(z_i)$ corresponds to the observational uncertainty related to the measured $H_\text{obs}(z_i)$.

\subsection{BAO Dataset}

Baryon Acoustic Oscillations (BAO) are periodic fluctuations in the density of baryonic matter of the universe, resulting from acoustic density waves that transported primordial plasma of the early universe. These oscillations provide a unique mark on the matter power spectrum, and the location of the BAO peaks yield significant cosmological insights.\\

BAO peak measurements allow the estimation of the Hubble distance $D_{H}(z)$ and the angular distance $D_{A}(z)$, enabling the constraint of parameters regarding DE. The acoustic horizon $r$ determined by the BAO peaks, acts as a standard ruler that permits us to figure out both the redshift separation and the angular separation at a certain redshift $z$, defined as: $\delta_z = r/D_{H}(z)$ and $\delta_\theta = r/(1+z)D_{A}(z)$. By adopting suitable asumptions for $r$ and constraining the cosmological parameters that determine the ratios $D_{A}(z)/r$ and $D_{H}(z)/r$, we may precisely calculate the Hubble parameter $H(z)$. This study employs 26 uncorrelated BAO radial measurements data points sourced from refs. \cite{10.1111/j.1365-2966.2009.15405.x,10.1093/mnras/stt1290,10.1093/mnras/stx721,10.1093/mnras/stu523,10.1093/mnras/stx1090,refId0,10.1093/mnras/stu111,10.1111/j.1365-2966.2012.21473.x,10.1093/mnras/stu523,osti_1407184,2017A&A...603A..12B}. The $\chi^2$ estimation for the BAO dataset is articulated as

\begin{equation}
    \chi^2_{B A O}(\phi) = \sum_{i=1}^{26} \left[ \frac{(H_{th}(z_i, \phi) - H_{obs}(z_i))^2}{\sigma^2_H(z_i)} \right],
\end{equation}

Here, $H_{th}(z_i, \phi)$ implies the theoretical Hubble parameter at redshift $z_i$ for a certain set of model parameters $\phi$, $H_{obs}(z_i)$ refers to the observed Hubble parameter derived from BAO observations, and $\sigma_H(z_i)$ pertains to the observational uncertainties associated with the BAO data.\\

\subsection{Pantheon+SH0ES Dataset}

Another baseline dataset employed in our MCMC analyses comprises observations from Type Ia supernova observations. These supernovae arise in binary star systems and play a key role in cosmology owing to their extremely uniform intrinsic luminosity, which lets them serve as reliable standard candles for the measurement of extragalactic distances. The distance modulus is defined as the difference between the measured apparent magnitude $m$ and the absolute magnitude $M$, which charecterizes the intrinsic luminosity of the source, and can be written as follows:

\begin{equation}
    \mu(z_k) = m - M = 5 \log_{10} [D_L(z_k,\psi)] + 25
\end{equation}

where $D_L(z_k,\psi)$ is the luminosity distance given by,

\begin{equation}
    D_L(z_k,\psi) = c (1 + z_k) \int_{0}^{z_k} \frac{dz'}{H(z')}
\end{equation}

Methodologically, the apparent magnitude of each Type Ia supernova is calibrated by adopting a fiducial absolute magnitude, $M$. Accordingly, within the MCMC framework, $M$ is regarded as a nuisance parameter and marginalized. Theoretical cosmological models are employed to estimate the distance modulus for a chosen parameter space, which are then compared with the supernova measurements compiled in the Pantheon datset. The cosmological parameters are constrained by minimizing a chi-square likelihood of the type \cite{Conley_2011},

\begin{equation}
    \chi^2_{\mathrm{SN}} = \Delta\mu^{T} C^{-1} \Delta\mu
\end{equation}

where $\Delta\mu = \mu_{\mathrm{th}}(z_k,\psi) - \mu_{\mathrm{obs}}(z_k)
$, and $C$ is the covariance matrix incorporating both statistical and systematic uncertainties. Equivalently, we can express it as,

\begin{equation}
    \chi^2(\psi) = \sum_{k=1}^{N} \sum_{l=1}^{N} 
    \big[ \mu_{\text{th}}(z_k,\psi) - \mu_{\text{obs}}(z_k) \big] 
    (C^{-1})_{kl} \\
    \times \big[ \mu_{\text{th}}(z_l,\psi) - \mu_{\text{obs}}(z_l) \big],
\end{equation}

This study employs the Pantheon+SH0ES\cite{Scolnic_2022} dataset of type Ia supernovae, which is an updated and expanded version of the original pantheon sample. The original Pantheon compilation had 1048 SNIa with a redshift range up to $z \sim 2.3$, while the Patheon+SH0ES collection has 1701 SNIa across an expanded redshift range $0.001<z<2.25$. The expanded coverage and addition of low-redshift supernovae facilitate a more rigorous analysis of systematic errors, providing enhanced precision in cosmological parameter constraints. 

Upon separately examining the $H(z)$, BAO, and Pantheon+SH0ES datasets, we establish the optimal values of the model through a joint investigation by combining all the datasets. To accomplish this, we define the aggregated $\chi^2$ function as the summation of the individual contributions from the observational Hubble parameter measurements $H(z)$, BAO, and Pantheon+SH0ES:

\begin{equation}
    \chi^2_{total}= \chi^2_{H(z)} + \chi^2_{BAO} + \chi^2_{SN}
\end{equation}

\begin{figure}[htbp]
    \centering
    \begin{minipage}{0.95\textwidth}
        \centering
        \begin{minipage}{0.45\textwidth}
            \centering
            \includegraphics[width=\linewidth]{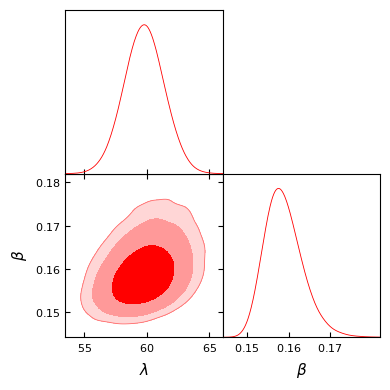}
            \vspace{0.2cm}
            \textbf{(i)}
        \end{minipage}
        \hfill
        \begin{minipage}{0.45\textwidth}
            \centering
            \includegraphics[width=\linewidth]{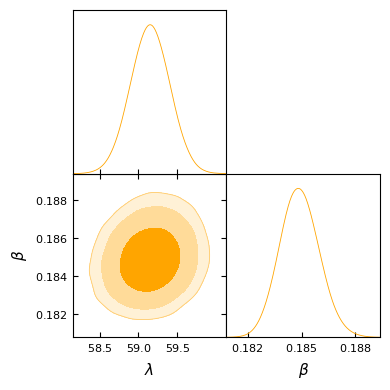}
            \vspace{0.2cm}
            \textbf{(ii)}
        \end{minipage}
        
        \vspace{0.5cm} 
        
        \begin{minipage}{0.45\textwidth}
            \centering
            \includegraphics[width=\linewidth]{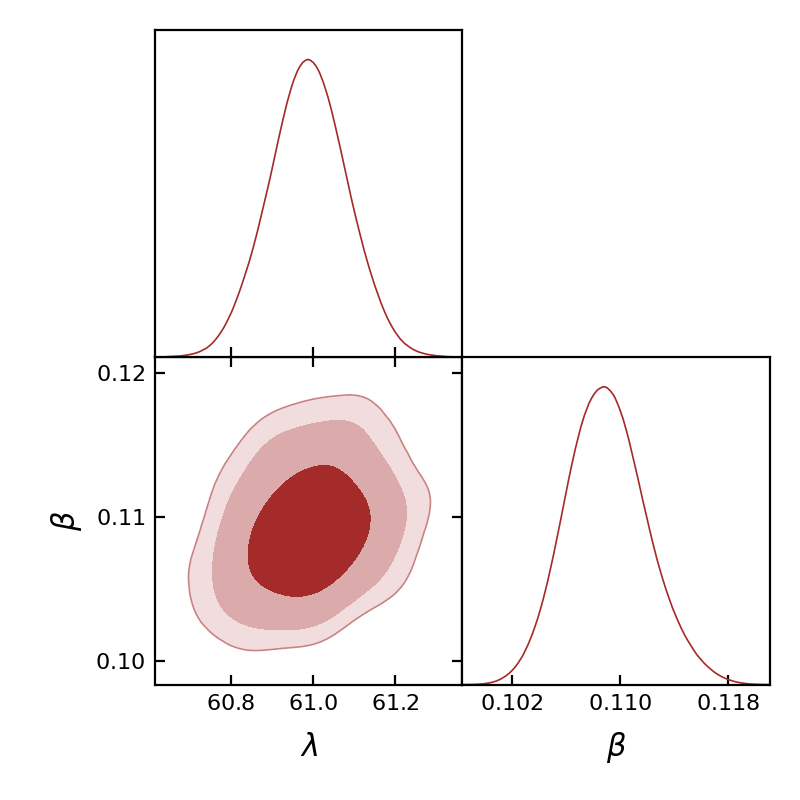}
            \vspace{0.2cm}
            \textbf{(iii)}
        \end{minipage}
        \hfill
        \begin{minipage}{0.45\textwidth}
            \centering
            \includegraphics[width=\linewidth]{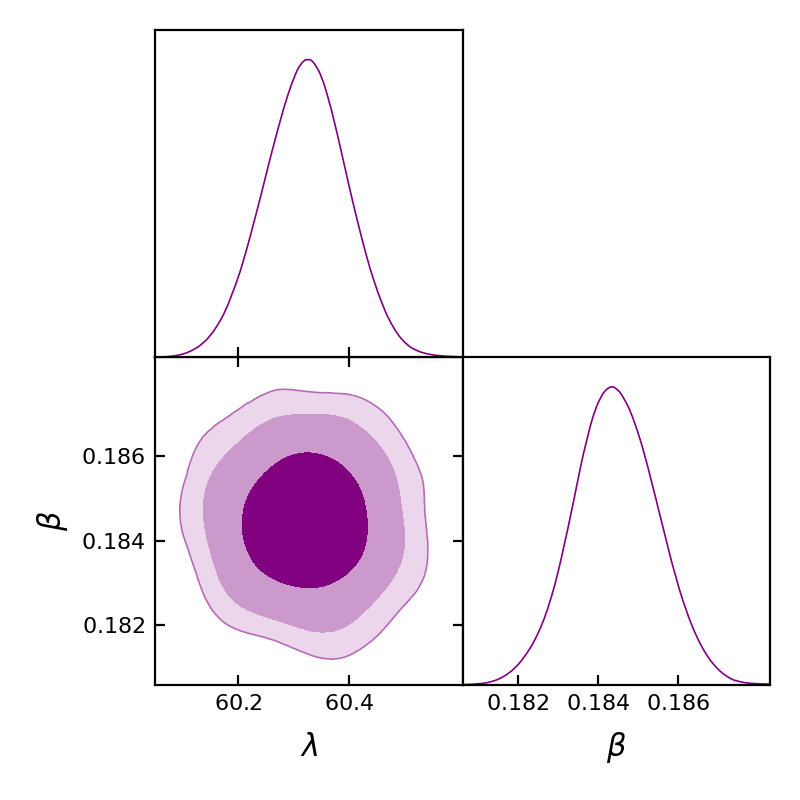}
            \vspace{0.2cm}
            \textbf{(iv)}
        \end{minipage}
    \end{minipage}
    
    \caption{2D-contour plots generated from the analysis of \textbf{(i)}~$H(z)$, 
    \textbf{(ii)}~BAO, \textbf{(iii)}~Pantheon+SH0ES, and \textbf{(iv)}~combined ($H(z)$ + BAO + Pantheon+SH0ES) datasets, 
    illustrating the best-fit values and up to $3\sigma$ confidence intervals for 
    the model parameters $\lambda$ and $\beta$.}
\end{figure}

\begin{table}[htbp]
\centering
\renewcommand{\arraystretch}{1.4} 
\setlength{\tabcolsep}{10pt} 
\captionsetup{justification=raggedright, singlelinecheck=false} 
\begin{tabular}{llccc}
\toprule
\textbf{Dataset} & & \boldmath{$\lambda$} & \boldmath{$\beta$}  \\ 
\hline
{Hubble} 
 & $1\sigma$ & $59.7734^{+1.6081}_{-1.6041}$ & $0.1584^{+0.005}_{-0.004}$  \\ 
 & $2\sigma$ & $59.7734^{+3.1744}_{-3.1774}$ & $0.1584^{+0.0108}_{-0.0074}$  \\ 
 & $3\sigma$ & $59.7734^{+4.7788}_{-4.7729}$ & $0.1584^{+0.0189}_{-0.0103}$  \\ 
\hline
{BAO} 
 & $1\sigma$ & $59.1499^{+0.2534}_{-0.2541}$ & $0.1849^{+0.0011}_{-0.0011}$  \\ 
 & $2\sigma$ & $59.1499^{+0.4970}_{-0.5009}$ & $0.1849^{+0.0022}_{-0.0021}$  \\ 
 & $3\sigma$ & $59.1499^{+0.7501}_{-0.7593}$ & $0.1849^{+0.0035}_{-0.0031}$  \\ 
\hline
{Pantheon+SH0ES} 
 & $1\sigma$ & $60.9872^{+0.0957}_{-0.0931}$ & $0.1090^{+0.0030}_{-0.0029}$  \\ 
 & $2\sigma$ & $60.9872^{+0.1877}_{-0.1911}$ & $0.1090^{+0.0063}_{-0.0056}$  \\ 
 & $3\sigma$ & $60.9872^{+0.2938}_{-0.2869}$ & $0.1090^{+0.0093}_{-0.0081}$  \\ 
\hline
{Combined} 
 & $1\sigma$ & $60.3234^{+0.0724}_{-0.0771}$ & $0.1844^{+0.0011}_{-0.0010}$  \\ 
 & $2\sigma$ & $60.3234^{+0.1437}_{-0.1509}$ & $0.1844^{+0.0021}_{-0.0021}$  \\ 
 & $3\sigma$ & $60.3234^{+0.2191}_{-0.2132}$ & $0.1844^{+0.0030}_{-0.0030}$  \\ 
\hline
\end{tabular}
\caption{Parameter constraints for $\lambda$ and $\beta$, exhibiting best-fit values with $1\sigma$, $2\sigma$, and $3\sigma$ confidence intervals derived from the $H(z)$, BAO, Pantheon+SH0ES and combined datasets.}
\end{table}

The \textbf{Fig.1} represents the corner plots of the model parameters $\lambda$ and $\beta$ up to $3\sigma$ confidence intervals extracted through MCMC approach by fitting the model with \textbf{(i)}$H(z)$, \textbf{(ii)}BAO, \textbf{(iii)}Pantheon+SH0ES and {(iv)}Combined datasets with the help of Bayesian statistics. In the \textbf{Table I}, the best-fit values of the model parameters with $1\sigma$ (68$\%$ confidence), $2\sigma$ (95$\%$ confidence), and $3\sigma$ (99$\%$) confidence intervals are presented and also, by considering these values we proceed further investigations. it is observed that the $1\sigma$ values of $\lambda$ and $\beta$ are varying within the range around $(59,60)$ and $(0.10,0.19)$, presenting their tight constraints and reflecting the robustness of the hybrid scale factor.\\

\begin{figure}[htbp]
    \centering
    \begin{subfigure}[t]{0.46\textwidth}
        \centering
        \includegraphics[width=\textwidth]{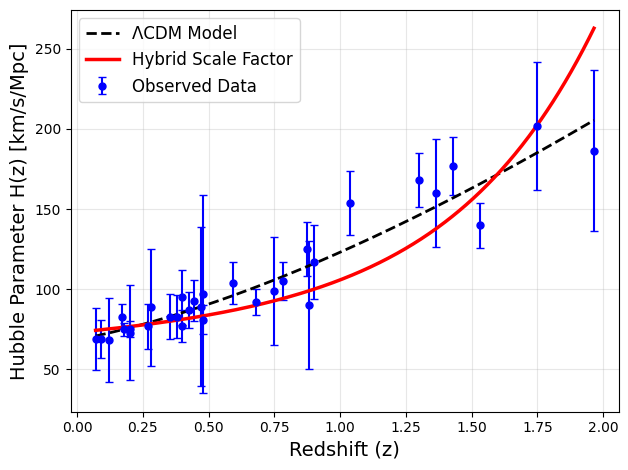}
    \end{subfigure}
    \hfill
    \begin{subfigure}[t]{0.48\textwidth}
        \centering
        \includegraphics[width=\textwidth]{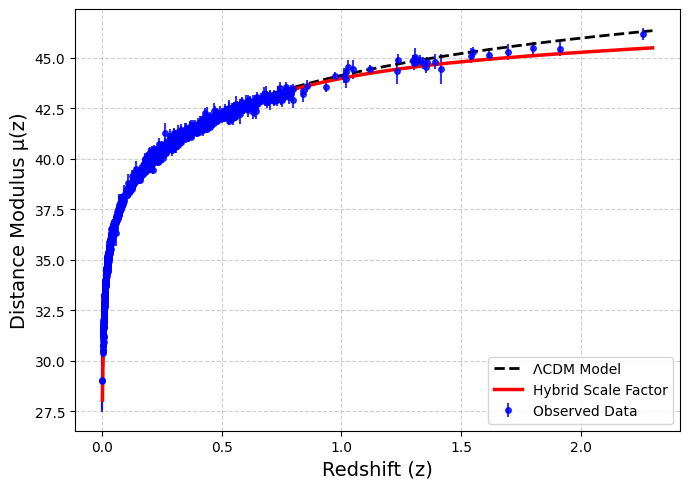}
    \end{subfigure}
    \caption{\raggedright The left panel illustrates the error bar plot of the Hubble parameter, whereas the right panel depicts the distance modulus function for the $\Lambda$CDM model and hybrid scale factor.}
\end{figure}

\begin{figure}[htbp]
    \centering
    \begin{subfigure}[t]{0.49\textwidth}
        \centering
        \includegraphics[width=\textwidth]{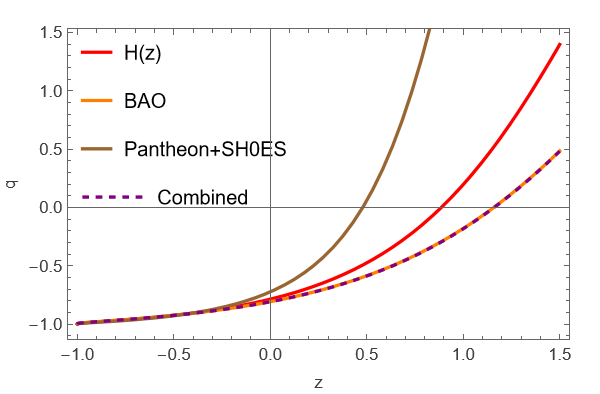}
    \end{subfigure}
    \caption{\raggedright Illustrates the plot of the deceleration parameter for $H(z)$, BAO, Pantheon+SH0ES, and combined datasets.}
\end{figure}

The left panel and right panel of \textbf{Fig.2} illustrates the error bar plot of the Hubble parameter and distance modulus function, respectively, showing that the Hubble model derived from the hybrid scale factor is greatly aligned with the $\Lambda$CDM model and observational data, with best-fit values of its parameters extracted through MCMC. The deceleration parameter presented in the \textbf{Fig.3} is staying negative throughout at present and in the late times for all the datasets, depicting the accelerating behaviour of the universe and shows transitions from positive values in the early universe to $q=-1$ at late times, denoting an evolution from a decelerated phase to an accelerated expansion for all considered datasets. Also, the values of $q$ at the redshift $z=0$ are $-0.79$, $-0.81$, $-0.73$, and $-0.81$ for $H(z)$, BAO, Pantheon+SH0ES and combined datasets and are in agreement with the literature \cite{Camarena_2020,doi:10.1142/S0217732322501048,Mandal_2023}.

\section{Dynamics in reconstructed $f(T)$ gravity models}

The primary objective of this study is to demonstrate that a class of theoretically motivated nonlinear $f(T)$ gravity models can provide a consistent description of the observed cosmic expansion when embedded within an observationally constrained framework based on the hybrid scale factor. In this context, the term \textit{reconstruction} refers not to the determination of an unknown gravitational action from observational data, but to the reconstruction of the cosmological dynamics, effective dark energy evolution, and observational viability of the proposed $f(T)$ models. This approach establishes the consistency of these models with the observed expansion history while preserving their theoretical motivation.

    Furthermore, before introducing the adopted $f(T)$ models, the Hubble parameter derived from the hybrid scale factor is used to obtain observational constraints on the cosmological expansion history. The viability of these non-linear $f(T)$ models is subsequently investigated through the evolution of cosmological parameters, behaviours of the equation of state parameter and energy conditions under the same observational background. 

\subsection{\textbf{Model I}: $f(T)=T-\frac{a}{(-T/6)^n}+b(\frac{-T}{6})^m$}
Here, we propose a type of $f(T)$ \cite{Yang_2011} theory analogous to the form of $f(R)=R-\alpha/R^n+\beta R^m$ \cite{PhysRevD.68.123512}, mentioned as,

\begin{equation}\label{eq:25}
    f(T)=T-\frac{a}{(-T/6)^n}+b(-T/6)^m
\end{equation}

which appeared to be the novel combination of two forms $\frac{a}{(-T/6)^n}$ \cite{PhysRevD.79.124019} and $b(\frac{-T}{6})^m$ \cite{PhysRevD.81.127301}. Furthermore, when $T\rightarrow \infty
$, GR is recovered if $m\leq1$ throughout early times; conversely, if $m>1$, this specific $f(T)$ theory diverges from GR even as $T \rightarrow \infty$. In later times, the term $-\frac{a}{(-T/6)^n}$ dominates and may cause accelerated expansion.  The model form in \eqref{eq:25} shows a linear behavior for $n=0$ and $m=1$.

Consecutively, the first and second derivatives of the $f(T)$ with respect to $T$, denoted as $f_{T}$ and $f_{TT}$ respectively, are determined and incorporated in equations  \eqref{eq:12} to \eqref{eq:14}.

As a result, the  equations reduced to;

\begin{equation}
    \rho_{de} = \frac{1}{2}{{\left( {{\left( \lambda +\frac{\lambda }{\zeta } \right)}^{2}} \right)}^{-n}}\left( 2an+a+b(2m-1){{\left( \frac{{{\lambda }^{2}}{{\left( \zeta +1 \right)}^{2}}}{{{\zeta }^{2}}} \right)}^{m+n}} \right),
\end{equation}

\begin{equation}
    p_{de} = \frac{{{\left( {{\left( \lambda +\frac{\lambda }{\zeta } \right)}^{2}} \right)}^{-n}}\left( b(2m-1)\left( -3\beta +2m-3\beta \zeta \left( \zeta +2 \right) \right){{\left( \frac{{{\lambda }^{2}}{{\left( \zeta +1 \right)}^{2}}}{{{\zeta }^{2}}} \right)}^{m+n}}-a(2n+1)\left( 3\beta +2n+3\beta \zeta \left( \zeta +2 \right) \right) \right)}{6\beta {{\left( \zeta +1 \right)}^{2}}},
\end{equation}

\begin{align}
\omega_{de} = -1-\frac{2an(2n+1)-2bm(2m-1){{\left( \frac{{{\lambda }^{2}}{{\left( \zeta +1 \right)}^{2}}}{{{\zeta }^{2}}} \right)}^{m+n}}}{3\beta {{\left( \zeta +1 \right)}^{2}}\left( 2an+a+b(2m-1){{\left( \frac{{{\lambda }^{2}}{{\left( \zeta +1 \right)}^{2}}}{{{\zeta }^{2}}} \right)}^{m+n}} \right)}
\end{align}

where,  $\zeta =W\left( \frac{\lambda {{\left( \frac{1}{z+1} \right)}^{1/\beta }}}{\beta } \right)$. Also, $\rho_{de}$, $p_{de}$, and $\omega_{de}$ represent the energy density, pressure, and EoS parameter for the DE sector.

\begin{figure}[htbp]
    \centering
    \begin{subfigure}[t]{0.49\textwidth}
        \centering
        \includegraphics[width=\textwidth]{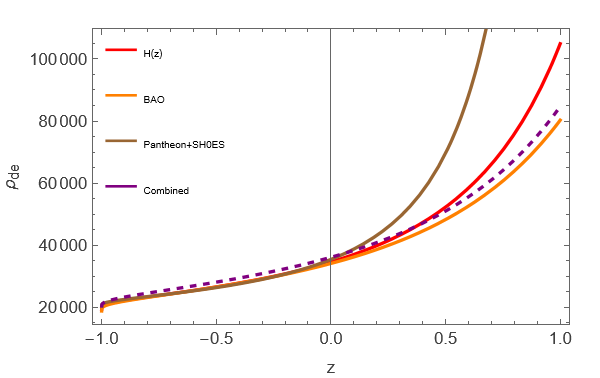}
        \caption*{(i) Energy density plot of Model I}
    \end{subfigure}
    \hfill
    \begin{subfigure}[t]{0.48\textwidth}
        \centering
        \includegraphics[width=\textwidth]{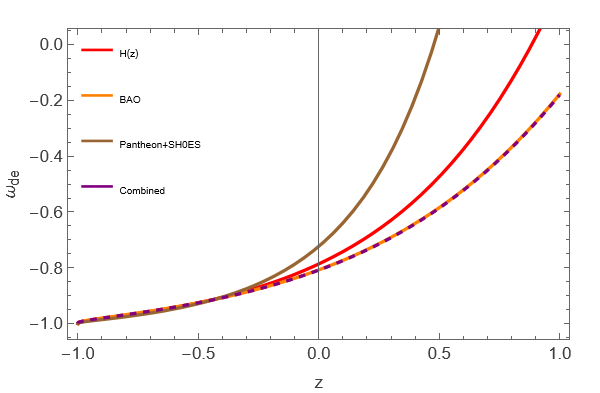}
        \caption*{(ii) EoS parameter plot of Model I}
    \end{subfigure}
    \caption{\raggedright The left panel displays energy density, and the right panel displays the EoS parameter plot of Model I for $H(z)$, BAO, Pantheon+SH0ES and combined datasets with the parameter scheme $a=0.665$, $b=0.084$,$m=1.505$, and $n=0.245$.}
\end{figure}

\textbf{Fig.4} illustrates the behaviours of energy density and EoS parameter of DE sector, depending upon the parameters $\lambda$, $\beta$, $m$, $n$, $a$, and $b$, where MCMC constrains the first two parameters, and the last three are selected to provide a positive energy density. Throughout the plot, we get negative EoS parameter and stays in the quintessence region at the present time and gives $-1$ at the late-time mimicking the $\Lambda$CDM behaviour for all observational datasets. The current values of the EoS parameter are around $\omega_0 =-0.78$, $\omega_0 = -0.808$, $\omega_0=-0.73$ and $\omega_0 = -0.809$ for $H(z)$, BAO, Pantheon+SH0ES, and combined datasets respectively which are consistent with the refs.\cite{Myrzakulov_2023,planck_collaboration}.

\subsection{\textbf{Model II}: $f(T)=T-aT_0 \left[\left(1+\frac{T^2}{T_0^2}\right)^{-n}-1 \right]$}
 
In analogous to a $f(R)$ model proposed by Satrobinsky, $f(R) = R+\alpha R_0\left[(1+\frac{R^2}{R_0^2})^{-n}-1\right]$\cite{Starobinsky_2007}, where $\alpha$,$R_0$, and $n$ are positive constants and $R_0$ is the order of the present Ricci scalar, here, we have proposed a form of $f(T)$\cite{Yang_2011}

\begin{equation}
    f(T)=T-aT_0 \left[\left(1+\frac{T^2}{T_0^2}\right)^{-n}-1 \right]
\end{equation}

where $a$, $n$ are positive constants, and $T_0$ is the order of the present Hubble parameter. This form reduces to a linear one in the absence of the constant $n$.\\

The rationale for this category of $f(T)$ theories is created from their $f(R)$ counterparts: firstly, the condition $f(0)=0$ ensures that in a flat spacetime, no cosmological constant appears, suggesting that the origin of DE is associated with geometry instead of a specific constant. Secondly, in both early and late cosmic epochs, such models emulate GR with a cosmological constant, but not a genuine constant; thereby, these theories can successfully undergo observational investigations and come out nearly indistinguishable from the $\Lambda$CDM model. Significantly, although $\Lambda$CDM struggles with the coincidence problem, these dynamic $f(T)$ models, due to their evolving behaviour, can function equivalently to a cosmological constant, bypassing the coincidence issue entirely, rendering them highly appealing for further research.\\

Incorporating, the first and second derivatives of this $f(T)$ form with respect to $T$ (denoted as $f_{T}$ and $f_{TT}$) in equations \eqref{eq:12} to \eqref{eq:14}, we get energy density, pressure and EoS parameter of the DE sector, respectively, as;

\begin{equation} \label{rho II}
    \rho_{de} =-\tau \left( 1-\frac{2an\tau {{\xi }^{-n-1}}}{T_0} \right)+\frac{1}{2}aT_0\left( {{\xi }^{-n}}-1 \right)+\tau ,
\end{equation}

\begin{equation} \label{pressure II}
\begin{aligned}
p_{de} =A-B+\frac{1}{2}aT_0\left( 1-{{\xi }^{-n}} \right)+C-\frac{6{{\lambda }^{2}}{{\left( \zeta +1 \right)}^{2}}}{{{\zeta }^{2}}}+\frac{2{{\lambda }^{2}}}{\beta {{\zeta }^{2}}}
\end{aligned}
\end{equation}

\begin{equation} \label{EoS II}
\begin{aligned}
\omega_{de} =-1-\frac{144a{{\lambda }^{4}}nT_0{{\left( \zeta +1 \right)}^{2}}{{\xi }^{-n}}D}{\beta E\left( -2\tau \left( 1-\frac{2an\tau {{\xi }^{-n-1}}}{T_0} \right)+aT_0\left( {{\xi }^{-n}}-1 \right)+2\tau  \right)}
\end{aligned}
\end{equation}

where the Greek letters  $\zeta =W\left( \frac{\lambda {{\left( \frac{1}{z+1} \right)}^{1/\beta }}}{\beta } \right)$, $\tau =6{{\left( \lambda +\frac{\lambda }{\zeta } \right)}^{2}}$, and $\xi =\left( \frac{{{\tau }^{2}}}{{{T_0}^{2}}}+1 \right)$ represent the common terms in the equations \ref{rho II},\ref{pressure II}, and \ref{EoS II} whereas the alphabets

\begin{equation}
   A=\frac{6{{\lambda }^{2}}{{\left( \zeta +1 \right)}^{2}}\left( T_0{{\zeta }^{2}}-12a{{\lambda }^{2}}n{{\left( \zeta +1 \right)}^{2}}{{\xi }^{-n-1}} \right)}{T_0{{\zeta }^{4}}}, 
\end{equation}

\begin{equation}
    B=\frac{2{{\lambda }^{2}}\left( T_0{{\zeta }^{2}}-12a{{\lambda }^{2}}n{{\left( \zeta +1 \right)}^{2}}{{\xi }^{-n-1}} \right)}{\beta T_0{{\zeta }^{4}}},
\end{equation}

\begin{equation}
    C=\frac{48a{{\lambda }^{4}}n{{\left( \zeta +1 \right)}^{2}}{{\xi }^{-n-2}}\left( -72{{\lambda }^{4}}(n+1){{\left( \zeta +1 \right)}^{4}}+{{T_0}^{2}}{{\zeta }^{4}}+36{{\lambda }^{4}}{{\left( \zeta +1 \right)}^{4}} \right)}{\beta {{T_0}^{3}}{{\zeta }^{8}}},
\end{equation}

\begin{equation}
    D=\left( 12{{\lambda }^{4}}(4n+1)+\left( 12{{\lambda }^{4}}(4n+1)-{{T_0}^{2}} \right){{\zeta }^{4}}+48{{\lambda }^{4}}(4n+1){{\zeta }^{3}}+72{{\lambda }^{4}}(4n+1){{\zeta }^{2}}+48{{\lambda }^{4}}(4n+1)\zeta  \right),  
\end{equation}

and 

\begin{equation}
   E={{\left( 36{{\lambda }^{4}}+\left( 36{{\lambda }^{4}}+{{T_0}^{2}} \right){{\zeta }^{4}}+144{{\lambda }^{4}}{{\zeta }^{3}}+216{{\lambda }^{4}}{{\zeta }^{2}}+144{{\lambda }^{4}}\zeta  \right)}^{2}} 
\end{equation}
 represent the some parts of those equations.

\begin{figure}[htbp]
    \centering
    \begin{subfigure}[t]{0.49\textwidth}
        \centering
        \includegraphics[width=\textwidth]{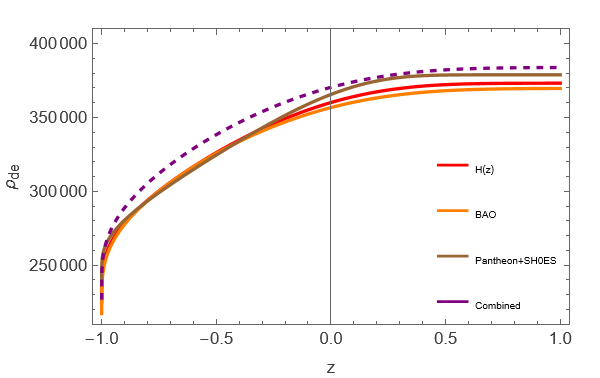}
        \caption*{(i) Energy density plot of model II}
    \end{subfigure}
    \hfill
    \begin{subfigure}[t]{0.48\textwidth}
        \centering
        \includegraphics[width=\textwidth]{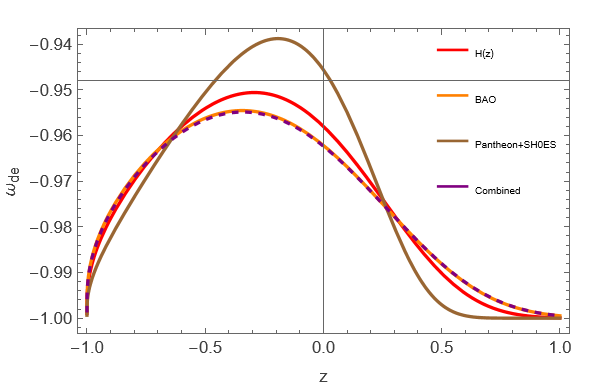}
        \caption*{(ii) EoS parameter plot of Model II}
    \end{subfigure}
    \caption{\raggedright The left panel displays energy density, and the right panel displays the EoS parameter plot of Model II for $H(z)$, BAO, Pantheon+SH0ES and combined datasets with the parameter scheme $a=23.2$ and $n=9.1$.}
\end{figure}

In the \textbf{Fig.5}, the parameters $\lambda$, $\beta$, $a$, and $n$ decide the behaviour of the energy density and the EoS parameter of the DE sector of which $\lambda$ and $\beta$ are constrained by the MCMC analysis, while $a$ and $n$ are chosen in such a manner to obtain the positive energy density. The EoS parameter stays negative throughout the plot indicating the quintessence phase of DE at the present time and it reaches to $-1$ at the late-time reflecting $\Lambda$CDM behaviour. While the current values of $\omega_{de}$ are around $-0.9581$, $-0.9624$, $-0.9457$,and $-0.9624$ for $H(z)$, BAO, Pantheon+SH0ES, and combined datasets, which are consistent with the literature \cite{planck_collaboration,doi:10.1142/S0217732322501048}. 

\subsection{\textbf{Model III}: $f(T)=T+aT_0\frac{(T^2/T_0^2)^n}{1+(T^2/T_0^2)^n}$}

In this subsection, a function can be considered as \cite{Yang_2011}

\begin{equation}
    f(T)=T+aT_0\frac{(T^2/T_0^2)^n}{1+(T^2/T_0^2)^n}
\end{equation}

This form is inspired by similar extensions in $f(R)$ gravity, $f(R)=R-\alpha R_0\frac{(R^2/R_0^2)^n}{1+(R^2/R_0^2)^n}$ \cite{PhysRevD.76.064004} that has been shown to align with both cosmological observations and local gravity tests. Also, this form reduces to linear form when the value of $n$ vanishes. \\

Moreover, the explanation for embracing this form in $f(T)$ gravity is threefold: Initially, at high redshift, the dynamics of the model closely align with the $\Lambda$CDM framework, offering it suitable for observational verification through cosmic microwave background (CMB) data. Secondly, at low redshift, it inherently triggers cosmic acceleration, demonstrating an expansion history akin to $\Lambda$CDM, but avoiding the necessity of a true cosmological constant. Third, it allows strictly regulated deviations from GR, incorporating $\Lambda$CDM as a specific instance, thereby enabling accurate cosmological constraints.

We, now, determine the derivatives of this $f(T)$ form such as $f_{T}$ and $f_{TT}$ in order to replace their values in equations \eqref{eq:12}
to \eqref{eq:14}.
Consequently, the reconstructed DE density, pressure and EoS parameter are obtained as;

\begin{equation}\label{DE density of form 3}
    \rho_{de} = \frac{1}{2}aT_0\left( \frac{{{36}^{n}}(4n+1)\eta +1}{{{\left( {{36}^{n}}\eta +1 \right)}^{2}}}-1 \right)
\end{equation}

\begin{equation} \label{DE pressure of model 3}
\begin{aligned}
p_{de} =\frac{a{{6}^{2n-1}}T_0\eta (F+G)}{\beta {{\left( \zeta +1 \right)}^{2}}{{\left( {{36}^{n}}\eta +1 \right)}^{3}}}
\end{aligned}
\end{equation}

\begin{equation} \label{DE EoS of model 3}
    \omega_{de} = -1-\frac{4n\left( \frac{1-4n}{{{36}^{n}}\eta -4n+1}-\frac{2}{{{36}^{n}}\eta +1} \right)}{3\beta {{\left( \zeta +1 \right)}^{2}}}
\end{equation}

where, as in model II, the Greek letters $\zeta =W\left( \frac{\lambda {{\left( \frac{1}{z+1} \right)}^{1/\beta }}}{\beta } \right)$, and $\eta ={{\left( \frac{{{\left( \lambda +\frac{\lambda }{\zeta } \right)}^{4}}}{{{T_0}^{2}}} \right)}^{n}}$ refer to the common terms in the equations \ref{DE density of form 3}, \ref{DE pressure of model 3}, and \ref{DE EoS of model 3} whereas the alphabets 

\begin{equation}
   F={{2}^{2n+1}}{{9}^{n}}\left( 3\beta -2n(3\beta +4n+1)-3\beta (2n-1)\zeta \left( \zeta +2 \right) \right)\eta , 
\end{equation}

and 

\begin{equation}
   G=\beta {{3}^{4n+1}}{{16}^{n}}{{\left( \zeta +1 \right)}^{2}}{{\eta }^{2}}+(4n-1)\left( -3\beta +4n-3\beta \zeta \left( \zeta +2 \right) \right)
\end{equation}

refer to some parts of those equations.

\begin{figure}[htbp]
    \centering
    \begin{subfigure}[t]{0.49\textwidth}
        \centering
        \includegraphics[width=\textwidth]{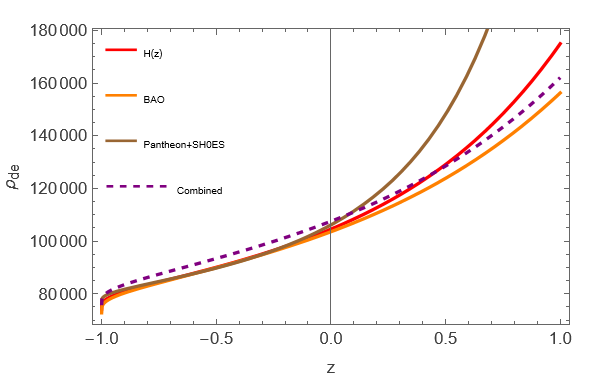}
        \caption*{(i) Energy density plot of model III}
    \end{subfigure}
    \hfill
    \begin{subfigure}[t]{0.48\textwidth}
        \centering
        \includegraphics[width=\textwidth]{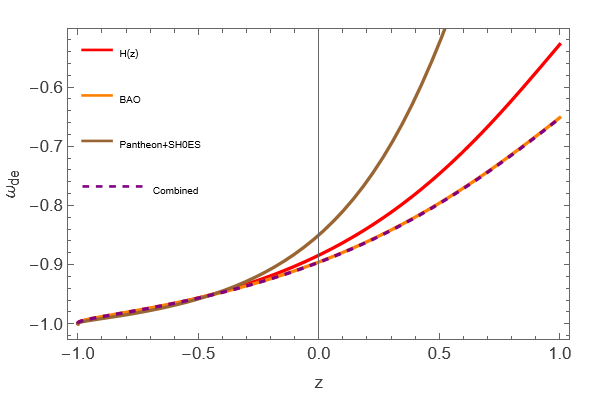}
        \caption*{(ii) EoS parameter plot of Model III}
    \end{subfigure}
    \caption{\raggedright The left panel displays energy density, and the right panel displays the EoS parameter plot of Model III for $H(z)$, BAO, Pantheon+SH0ES and combined datasets with the parameter scheme $a=34.2$ and $n=0.31$.}
\end{figure}

\vspace{0.5cm}

The expressions present in equations \eqref{DE density of form 3} and \eqref{DE EoS of model 3} can be graphically analysed for different datasets
because they are dependent upon the Hubble parameter derived from the hybrid scale factor. We can see \textbf{Fig.6} that the energy density of the DE sector for Model III remains positive throughout, driven by the parameters $\lambda$, $\beta$, $m$, and $n$ for the three data sets, satisfying the accelerated expansion nature of the universe. Also, the EoS parameter remains negative for all the datasets reflecting quintessence behaviour at the present time and moving towards $-1$ and giving $\Lambda$CDM behaviour at the late-time. We can follow the literature \cite{Sudharani:2023,Koussour:2024} with which the current values of the EoS parameter around $\omega_0=-0.885$, $\omega_0=-0.895$, $\omega_0=-0.848$, and $\omega_0=-0.896$  for $H(z)$, BAO, Pantheon+SH0ES, and combined datasets, respectively, are consistent.

\vspace{1cm}
Based on the results of three above functional forms, a comparison table can be given as

\begin{table}[htbp]
\centering
\renewcommand{\arraystretch}{1.2}

\resizebox{\textwidth}{!}{%
\begin{tabular}{|c|c|c|c|c|c|c|c|c|c|c|c|c|}
\hline
\multirow{2}{*}{\textbf{Parameter}} &
\multicolumn{4}{|c|}{\textbf{Model I}} &
\multicolumn{4}{|c|}{\textbf{Model II}} &
\multicolumn{4}{|c|}{\textbf{Model III}} \\ \cline{2-13}
 & \textbf{H(z)} & \textbf{BAO} & \textbf{Pantheon+SH0ES} & \textbf{Combined} 
 & \textbf{H(z)} & \textbf{BAO} & \textbf{Pantheon+SH0ES} & \textbf{Combined}
 & \textbf{H(z)} & \textbf{BAO} & \textbf{Pantheon+SH0ES} & \textbf{Combined} \\ \hline
$H(z_0)\,{\rm Km\,s^{-1}\,Mpc^{-1}}$
 & 73.2296 & 72.8813 & 73.7845 &  74.2685
 & 73.2296 & 72.8813 & 73.7845 &  74.2685
 & 73.2296 & 72.8813 & 73.7845 & 74.2685 \\ \hline
$q(z_0)$
 & -0.79 & -0.809 & -0.73 & -0.805 
 & -0.79 & -0.809 & -0.73 & -0.805 
 & -0.79 & -0.809 & -0.73 & -0.805 \\ \hline
$\omega(z_0)$
 & -0.78 & -0.808 & -0.73 & -0.809 
 & -0.9581 & -0.9624 & -0.9457 & -0.9624 
 & -0.885 & -0.895 & -0.848 & -0.896 \\ \hline
\end{tabular}%
}

\caption{Comparison of parameters for Model I, Model II, and Model III using different datasets.}
\end{table}

The results presented in the table indicate that, at the present epoch (z=0), the equation-of-state (EoS) parameter for all three reconstructed models resides within the quintessence regime, regardless of their distinct evolutionary trajectories at other redshifts. Furthermore, the present-day values of the Hubble parameter and the deceleration parameter are independent of the specific model construction. Consequently, these quantities attain identical values across all three proposed $f(T)$ formulations at zero redshift.

\section{Energy Conditions}

For further analysis, in this section, we investigate the behaviour of the energy conditions. The energy conditions are essential concepts for understanding the behaviour of geodesics in the universe, derivable from the Raychaudhuri equations and can be represented as

\begin{itemize}
    \item \textbf{Null Energy Condition (NEC):} $\rho + p \geq 0$;
    \item \textbf{Weak Energy Condition (WEC):} $\rho \geq 0$ and $\rho + p \geq 0$;
    \item \textbf{Strong Energy Condition (SEC):} $\rho + 3p \geq 0$;
    \item \textbf{Dominant Energy Condition (DEC):} $\rho - p \geq 0$
\end{itemize}

Examining energy conditions enables for an understanding of the characteristics of matter and energy in the universe, an essential step in comprehending its accelerated expansion and the impact of DE. Energy conditions fundamentally serve as boundary restrictions that regulate the evolution of the cosmos \cite{PhysRevD.68.023509}. Moreover, due to the fundamental causal structure of spacetime, gravitational interactions are fundamentally shaped by energy conditions \cite{doi:10.1142/S0218271819300167}. 

\begin{figure}[htbp]
    \centering
    \framebox{
        \begin{minipage}{\textwidth}
            \centering
            \begin{minipage}{0.32\textwidth}
                \centering
                \includegraphics[width=\textwidth]{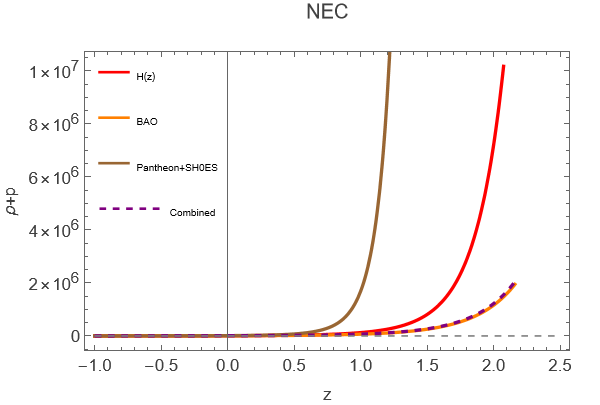}
            \end{minipage}
            \hfill
            \begin{minipage}{0.32\textwidth}
                \centering
                \includegraphics[width=\textwidth]{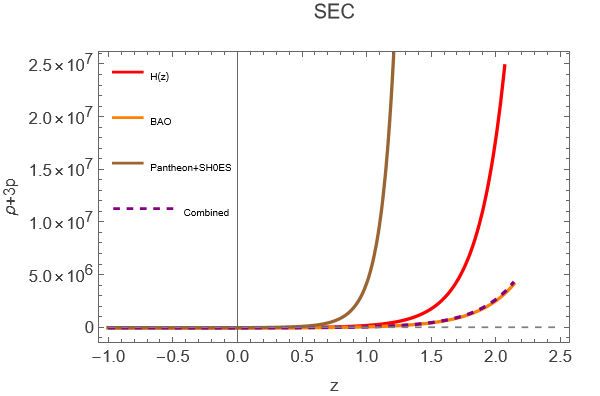}
            \end{minipage}
            \hfill
            \begin{minipage}{0.32\textwidth}
                \centering
                \includegraphics[width=\textwidth]{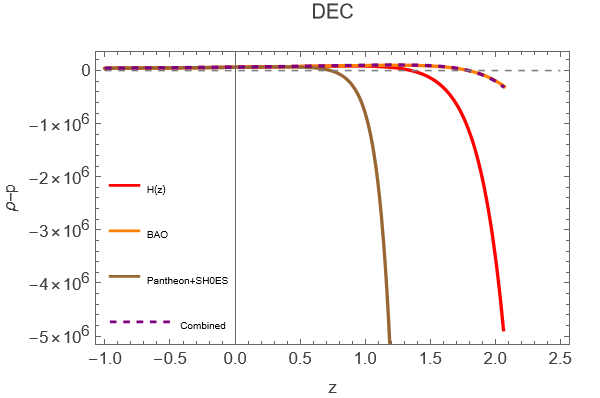}
            \end{minipage}
        \end{minipage}
    }
    \captionsetup{justification=raggedright, singlelinecheck=false}
    \caption{Energy conditions versus redshift for Model I over $H(z)$, BAO, Pantheon+SH0ES and combined datasets with the parameter scheme $a=0.665$, $b=0.084$,$m=1.505$, and $n=0.245$.}
\end{figure}

\begin{figure}[htbp]
    \centering
    \framebox{
        \begin{minipage}{\textwidth}
            \centering
            \begin{minipage}{0.32\textwidth}
                \centering
                \includegraphics[width=\textwidth]{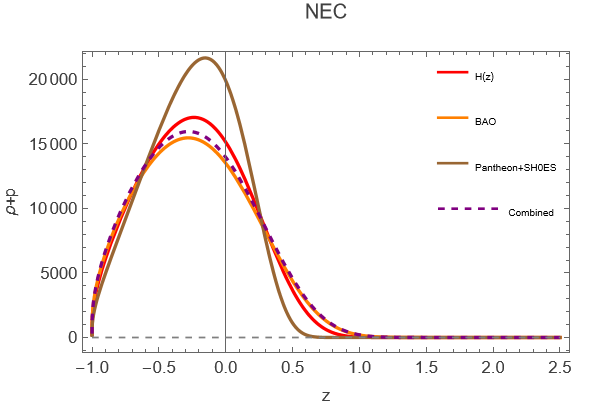}
            \end{minipage}
            \hfill
            \begin{minipage}{0.32\textwidth}
                \centering
                \includegraphics[width=\textwidth]{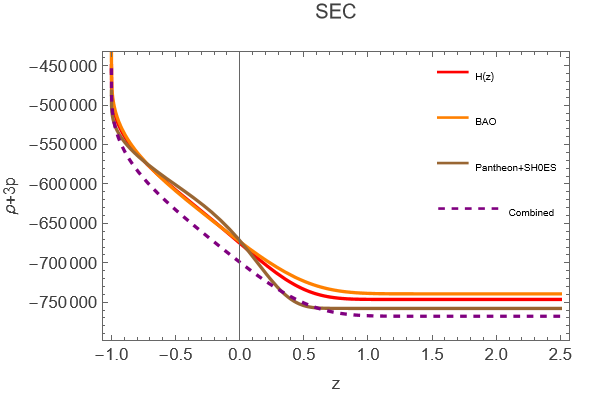}
            \end{minipage}
            \hfill
            \begin{minipage}{0.32\textwidth}
                \centering
                \includegraphics[width=\textwidth]{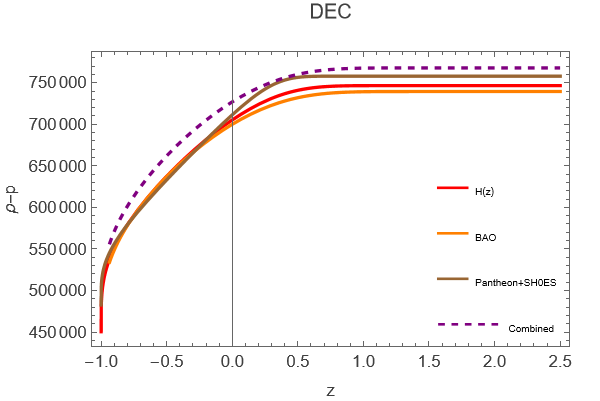}
            \end{minipage}
        \end{minipage}
    }
    \captionsetup{justification=raggedright, singlelinecheck=false}
    \caption{Energy conditions versus redshift for Model II over $H(z)$, BAO, Pantheon+SH0ES and combined datasets with the parameter scheme $a=23.2$ and $n=9.1$.}
\end{figure}

\begin{figure}[htbp]
    \centering
    \framebox{
        \begin{minipage}{\textwidth}
            \centering
            \begin{minipage}{0.32\textwidth}
                \centering
                \includegraphics[width=\textwidth]{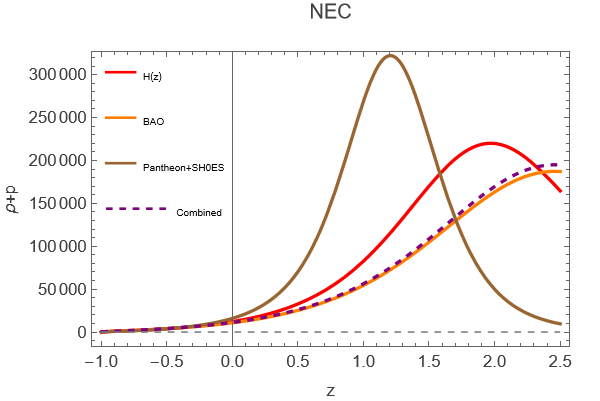}
            \end{minipage}
            \hfill
            \begin{minipage}{0.32\textwidth}
                \centering
                \includegraphics[width=\textwidth]{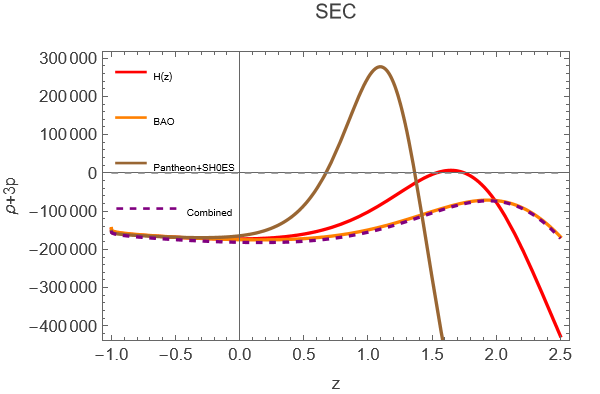}
            \end{minipage}
            \hfill
            \begin{minipage}{0.32\textwidth}
                \centering
                \includegraphics[width=\textwidth]{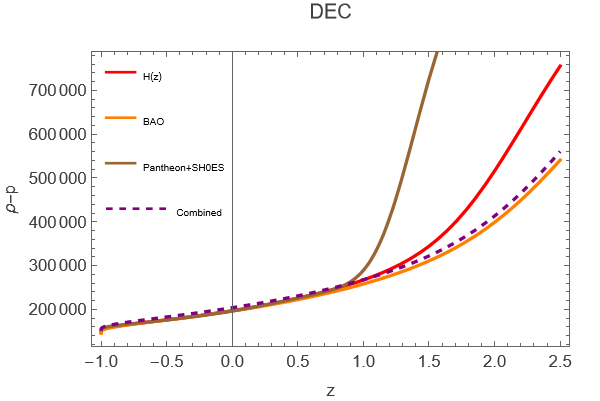}
            \end{minipage}
        \end{minipage}
    }
    \captionsetup{justification=raggedright, singlelinecheck=false}
    \caption{Energy conditions versus redshift for Model III over $H(z)$, BAO, Pantheon+SH0ES and combined datasets with the parameter scheme $a=34.2$ and $n=0.31$.}
\end{figure}

\textbf{Fig.7} exhibits the behaviour of the energy conditions for Model I for $H(z)$, BAO, Pantheon+SH0ES and combination of all. For all three datset, NEC is getting almost satisfied everywhere while DEC is getting satisfied for the redshift range $z \in [-1,1.34)$, $z \in [-1,1.79)$, $z \in [-1,0.73)$, and $z \in [-1,1.79)$ for $H(z)$, BAO, Pantheon+SH0ES, and combined datasets respectively. But SEC is getting violated in the redshift range around $[-1,0.63)$, $[-1,0.84)$, $[-1,0.32)$, and $[-1,0.84)$ for $H(z)$, BAO, Pantheon+SH0ES and combined datasets, respectively, which reflects the accelerated expansion at the current epoch and late-time. The energy conditions of Model II in \textbf{Fig.8} illustrates that the NEC and DEC are satisfied everywhere while SEC is getting violeted throughout the plot reflecting the expansion of the universe in an accelerated manner. The behaviour of the energy conditions in \textbf{Fig.9} demonstrates the accelerated expansion of the cosmos exhibiting the violating behaviour of SEC at the present and late-time, while it satisfies for the datasets $H(z)$ and Pantheon+SH0ES in the redshift range around $z \in (1.54,1.74)$ and $0.68,1.36$ respectively, while NEC and DEC are getting satisfied everywhere.

\section{Conclusion}

In the present work, we reconstruct $f(T)$ gravity by examining three distinct functional forms of $f(T)$ and solving the corresponding modified field equations within the Friedmann–Lema$\hat{i}$tre–Robertson–Walker (FLRW) cosmological framework. The cosmic expansion is modeled through a hybrid scale factor that incorporates both power-law and exponential behaviors. The free parameters associated with this hybrid parametrization are constrained using a Markov Chain Monte Carlo (MCMC) analysis under the framework of Bayesian Statistics and applied to multiple observational datasets, such as; $H(z),$ BAO, Pantheon+SH0ES, and their joint compilation. By confronting the theoretical predictions with observational evidence, we obtain bounds on the model parameters and verify that the reconstructed scenarios provide an adequate description of the Universe’s dynamical evolution. This approach contributes to a deeper understanding of the expansion history and underlying properties of the cosmos.\\

This analysis leads to several noteworthy results,
\begin{itemize}
    \item 
    The proposed hybrid scale factor exhibits strong agreement with the standard $\Lambda$CDM scenario and observational data, as evidenced by the Hubble and distance modulus analyses. By combining power-law and exponential terms, it successfully captures the transition from an early inflationary regime to a late-time accelerated phase. The present epoch behavior closely mimics that of the concordance $\Lambda$CDM model. Observational indications of late-time acceleration imply a transition from an earlier decelerating phase to the current accelerating phase, characterized by a signature change in the deceleration parameter. Unlike pure power law or exponential scale factors, which typically yield constant deceleration parameters and are therefore less suitable for modeling cosmic speed-up, the hybrid form effectively reproduces a dynamically evolving deceleration parameter with positive values at early times and negative values at late times. Hence, the resulting Hubble and deceleration parameters reflect behaviour consistent with observational cosmology, such as late-time $\Lambda$CDM-like evolution and the transition from deceleration to acceleration.

    \item The deceleration parameter, for all considered datasets, stays negative at present and late-time but shows a transition from deceleration to acceleration phase in the early universe. The transition from deceleration to acceleration for Hubble, BAO, Pantheon+SH0ES and combined dataset shows the value $z_t = 0.89,$ $z_t = 1.16,$ $z_t = 0.48$ and $z_t = 1.16$ respectively. The present value of the deceleration parameter has been obtained as $q_0 = -0.79,$ $q_0 = -0.809,$ $q_0 = -0.73$ and $q_0 = -0.805$ in the similar fashion which indicates the current accelerating nature of the universe \cite{doi:10.1142/S0217732322501048,Mandal_2023}.

    \item The energy density of the DE sector for all three models remains positive throughout the plot, ensuring the physical viability of the reconstructed scenarios.

    \item According to the results presented in Table II, the equation-of-state (EoS) parameter for all models evolves in a quintessence-like regime for all datasets. Moreover, the current values of the EoS parameter are in good agreement with previously reported results in the literature \cite{Myrzakulov_2023,planck_collaboration,doi:10.1142/S0217732322501048,Sudharani:2023,Koussour:2024}.

    \item 
    To assess the viability of the modified gravity framework, the standard energy conditions; namely, the null energy condition (NEC) and dominant energy condition (DEC) are examined. Depending on the redshift interval and model considered, these conditions may be either satisfied or violated. The strong energy condition (SEC) is violated near the present epoch and in the late-time for Model I and III and throughout the evolution for Model II across all datasets.
    
\end{itemize}

In conclusion, the present work provides a comprehensive analysis of late-time cosmological evolution within the framework of non-linear $f(T)$ gravity. The key geometrical and dynamical quantities are estimated using a Bayesian statistical framework, ensuring robust parameter inference. For all three reconstructed models, the obtained parameter values exhibit strong consistency with recent high-precision satellite observations. Future investigations,particularly those involving dynamical stability assessments and the inclusion of more recent observational datasets, would further enhance the reliability and scope of the proposed models.

\section*{References}

\bibliographystyle{JCAP.bst}
\bibliography{references}

\end{document}